\documentclass[prl,aps,twocolumn,floats,nofootinbib,superscriptaddress]{revtex4-1}
\usepackage{amsmath,amssymb,graphicx,psfrag}
\usepackage[colorlinks,linkcolor={blue},citecolor={blue},urlcolor={blue}]{hyperref}

\begin{document}
\renewcommand{\ni}{{\noindent}}
\newcommand{\dprime}{{\prime\prime}}
\newcommand{\be}{\begin{equation}}
\newcommand{\ee}{\end{equation}}
\newcommand{\bea}{\begin{eqnarray}} 
\newcommand{\eea}{\end{eqnarray}}
\newcommand{\la}{\langle}
\newcommand{\ra}{\rangle} 

\newcommand{\dg}{\dagger}
\newcommand\lbs{\left[}
\newcommand\rbs{\right]}
\newcommand\lbr{\left(}
\newcommand\rbr{\right)}
\newcommand\f{\frac}
\newcommand\e{\epsilon}
\newcommand\ua{\uparrow}
\newcommand\da{\downarrow}
\newcommand{\bcen}{\begin{center}}
\newcommand{\ecen}{\end{center}}
\newcommand{\btab}{\begin{tabular}}
\newcommand{\etab}{\end{tabular}}
\newcommand{\bdes}{\begin{description}}
\newcommand{\edes}{\end{description}}
\newcommand{\mc}{\multicolumn}
\newcommand{\ul}{\underline}
\newcommand{\non}{\nonumber}
\newcommand{\etal}{et.~al.\ }
\newcommand{\half}{\frac{1}{2}}
\newcommand{\bary}{\begin{array}}
\newcommand{\eary}{\end{array}}
\newcommand{\benum}{\begin{enumerate}}
\newcommand{\eenum}{\end{enumerate}}
\newcommand{\bitem}{\begin{itemize}}
\newcommand{\eitem}{\end{itemize}}
\newcommand{\cuup}[1]{c_{#1 \uparrow}}
\newcommand{\cdown}[1]{c_{#1 \downarrow}}
\newcommand{\cdup}[1]{c^\dagger_{#1 \uparrow}}
\newcommand{\cddown}[1]{c^\dagger_{#1 \downarrow}}
%
%
\newcommand{\beps}{\mbox{\boldmath $ \epsilon $}}
\newcommand{\bsig}{\mbox{\boldmath $ \sigma $}}
\newcommand{\bpi}{\mbox{\boldmath $ \pi $}}
\newcommand{\bkap}{\mbox{\boldmath $ \kappa $}}
\newcommand{\bgam}{\mbox{\boldmath $ \gamma $}}
\newcommand{\bphi}{\mbox{\boldmath $ \phi $}}
\newcommand{\balp}{\mbox{\boldmath $ \alpha $}}
\newcommand{\beot}{\mbox{\boldmath $ \eta $}}
\newcommand{\btau}{\mbox{\boldmath $ \tau $}}
\newcommand{\blam}{\mbox{\boldmath $ \lambda $}}
\newcommand{\bomg}{\mbox{\boldmath $ \omega $}}
\newcommand{\bOmg}{\mbox{\boldmath $ \Omega $}}
\newcommand{\bxhi}{\mbox{\boldmath $ \xi $}}
\newcommand{\bmu} {\mbox{\boldmath $ \mu $}}
\newcommand{\bnu} {\mbox{\boldmath $ \nu $}}
\newcommand{\bdelta}{{\boldsymbol{\delta}}}
\newcommand{\bTheta}{{\boldsymbol{\Theta}}}
\newcommand{\bpsi}{\mbox{\boldmath $ \psi $}}
\newcommand{\brho}{\mbox{\boldmath $ \rho $}}
\newcommand{\bGam}{\mbox{\boldmath $ \Gamma $}}
\newcommand{\bLam}{\mbox{\boldmath $ \Lambda $}}
\newcommand{\bPhi}{\mbox{\boldmath $ \Phi $}}
%
%
\newcommand{\ba} { \bm{a} }
\newcommand{\bb} { \mbox{\boldmath $b$}}
\newcommand{\bc} { {\mathbf c} }
\newcommand{\bd} { \mbox{\boldmath $d$}}
\newcommand{\bff}{ \mbox{\boldmath $f$}}
\newcommand{\bg} { \mbox{\boldmath $g$}}
\newcommand{\bh} { \mbox{\boldmath $h$}}
\newcommand{\bi} { \mbox{\boldmath $i$}}
\newcommand{\bj} { \mbox{\boldmath $j$}}
\newcommand{\bk} { \bm{k} }
\newcommand{\bl} { \mbox{\boldmath $l$}} 
\newcommand{\bmm} { \mbox{\boldmath $m$}}
\newcommand{\bn} { \mbox{\boldmath $n$}}
\newcommand{\bo} { \mbox{\boldmath $o$}}
\newcommand{\bp} { \bm{p} }
\newcommand{\bq} { \bm{q} }
\newcommand{\br} { \boldsymbol{r}}
\newcommand{\bs} { \mbox{\boldmath $s$}}
\newcommand{\bt} {\boldsymbol{t}} 
\newcommand{\bu} { \mbox{\boldmath $u$}}
\newcommand{\bv} { \mbox{\boldmath $v$}}
\newcommand{\bw} { \mbox{\boldmath $w$}}
\newcommand{\bx} { \mbox{\boldmath $x$}}
\newcommand{\by} { \mbox{\boldmath $y$}}
\newcommand{\bz} { \mbox{\boldmath $z$}}
\newcommand{\bA} { \mbox{\boldmath $A$}}
\newcommand{\bB} { \mbox{\boldmath $B$}}
\newcommand{\bC} { \mbox{\boldmath $C$}}
\newcommand{\bD} { \mbox{\boldmath $D$}}
\newcommand{\bF} { \mbox{\boldmath $F$}}
\newcommand{\bG} { \mbox{\boldmath $G$}}
\newcommand{\bH} { \mbox{\boldmath $H$}}
\newcommand{\bI} { \mbox{\boldmath $I$}}
\newcommand{\bJ} { \mbox{\boldmath $J$}}
\newcommand{\bK} { \mbox{\boldmath $K$}}
\newcommand{\bL} { \mbox{\boldmath $L$}}
\newcommand{\bM} { \mbox{\boldmath $M$}}
\newcommand{\bN} { \mbox{\boldmath $N$}}
\newcommand{\bO} { \mbox{\boldmath $O$}}
\newcommand{\bP} { \mbox{\boldmath $P$}}
\newcommand{\bQ} { \boldsymbol{Q} }
\newcommand{\bR} { {\mathbf R} }
\newcommand{\bS} { \mbox{\boldmath $S$}}
\newcommand{\bT} { \mbox{\boldmath $T$}}
\newcommand{\bU} { \mbox{\boldmath $U$}}
\newcommand{\bV} { \mbox{\boldmath $V$}}
\newcommand{\bW} { \mbox{\boldmath $W$}}
\newcommand{\bX} { \mbox{\boldmath $X$}}
\newcommand{\bY} { \mbox{\boldmath $Y$}}
\newcommand{\bZ} { \mbox{\boldmath $Z$}}
\newcommand{\bzero} { \mbox{\boldmath $0$}}
\newcommand{\bfell} {\mbox{\boldmath $ \ell $}}

%
%
\newcommand{\dou}{\partial}
\newcommand{\leftjb} {[\![}
\newcommand{\rightjb} {]\!]}
\newcommand{\ju}[1]{ \leftjb #1 \rightjb }
\newcommand{\D}[1]{\mbox{d}{#1}} 
\newcommand{\grad}{\mbox{\boldmath $\nabla$}}
\newcommand{\modulus}[1]{|#1|}
\renewcommand{\div}[1]{\grad \cdot #1}
\newcommand{\curl}[1]{\grad \times #1}
\newcommand{\mean}[1]{\langle #1 \rangle}
\newcommand{\bra}[1]{{\langle #1 |}}
\newcommand{\ket}[1]{| #1 \rangle}
\newcommand{\braket}[2]{\langle #1 | #2 \rangle}
\newcommand{\dbdou}[2]{\frac{\dou #1}{\dou #2}}
\newcommand{\dbdsq}[2]{\frac{\dou^2 #1}{\dou #2^2}}
\newcommand{\Pint}[2]{ P \!\!\!\!\!\!\!\int_{#1}^{#2}}
\newcommand{\Itwo}{{\mathds{1}}}
\newcommand{\Hds}{{\mathds{H}}}
\newcommand{\cH}{{\cal H}}
\newcommand{\cS}{{\cal S}}

%
%
\newcommand{\prn}[1] {(\ref{#1})}
\newcommand{\sect}[1] {Section~\ref{#1}}
\newcommand{\Sect}[1] {Section~\ref{#1}}

%
%
\newcommand{\uncon}[1]{\centerline{\epsfysize=#1 \epsfbox{/usr2/yogeshwar/styles/construction.pdf}}}
\newcommand{\checkup}[1]{{(\tt #1)}\typeout{#1}}
\newcommand{\ttd}[1]{{\color[rgb]{1,0,0}{\bf #1}}}
\newcommand{\red}[1]{{\color[rgb]{1,0,0}{\protect{#1}}}}
\newcommand{\blue}[1]{{\color[rgb]{0,0,1}{#1}}}
\newcommand{\green}[1]{{\color[rgb]{0.0,0.5,0.0}{#1}}}
\newcommand{\citebyname}[1]{\citeauthor{#1}\cite{#1}}
\newcommand{\myfigwidth}{0.95\columnwidth}
\newcommand{\myhalffig}{0.475\columnwidth}
\newcommand{\mythirdfig}{0.33\columnwidth}
\newcommand{\signum}[0]{\mathop{\mathrm{sign}}}
\newcommand{\skup}{\ket{s \uparrow}}
\newcommand{\skdn}{\ket{s \downarrow}}
\newcommand{\pkup}{\ket{p \uparrow}}
\newcommand{\pkdn}{\ket{p \downarrow}}
\newcommand{\sbup}{\bra{s \uparrow}}
\newcommand{\sbdn}{\bra{s \downarrow}}
\newcommand{\pbup}{\bra{p \uparrow}}
\newcommand{\pbdn}{\bra{p \downarrow}}

\newcommand{\Eqn}[1] {Eqn.~(\ref{#1})}
\newcommand{\Fig}[1]{Fig.~\ref{#1}}

\title{Single-particle excitations across the localization and many-body localization transition in quasi-periodic systems}
\author{Yogeshwar Prasad}
\affiliation{Department of Liberal Studies, Kangwon National University, Samcheok, 25913, Republic of Korea}
\affiliation{Theory Division, Saha Institute of Nuclear Physics, 1/AF Bidhannagar, Kolkata 700 064, India}
\author{Arti Garg}
\affiliation{Theory Division, Saha Institute of Nuclear Physics, 1/AF Bidhannagar, Kolkata 700 064, India}
 \affiliation{Homi Bhabha National Institute, Training School Complex, Anushaktinagar, Mumbai 400094, India}
\vspace{0.2cm}
\begin{abstract}
\vspace{0.3cm}
       {We study localization and many-body localization transition in one dimensional systems in the presence of deterministic quasi-periodic potential. We use single-particle excitations obtained through single-particle Green's function in real space to characterize the localization to delocalization transition. A single parameter scaling analysis of the ratio of the typical to average value of the local density of states (LDOS) of single particle excitations shows that the critical exponent with which the correlation length $\xi$ diverges at the transition point $\xi \sim |h-h_c|^{-\nu}$, coming from the localized side, satisfies the inequality $\nu \ge 1$ for the non-interacting Aubry-Andre (AA) model. For the interacting system with AA potential, we study single particle excitations produced in highly excited many-body eigenstates across the MBL transition and found that the critical exponent obtained from finite-size scaling of the ratio of the typical to average value of the LDOS satisfies $\nu \ge 1$ here as well. This analysis of local density of states shows that the localization and MBL transition in systems with quasi-periodic potential belong to a different universality class than the localization and MBL transition in systems with random disorder where $\nu \ge 2$. In complete contrast to this, finite-size scaling of the level spacing ratio is known to support the same universality class for MBL transitions in systems with quasiperiodic as well as random disorder potentials. For the interacting systems with quasiperiodic potentials, though finite-size scaling of the level spacing ratio shows a transition at $h_c^{lsr}$ which is close to the transition point obtained from LDOS within numerical precision, the critical exponent obtained from finite-size scaling of level spacing ratio is $\nu \sim 0.54$ in close similarity to the MBL systems with random disorder.}
        
\end{abstract} 
\maketitle
\section{I. Introduction}
Transport properties of a system are significantly influenced by disorder. Any tiny quantity of disorder is sufficient to localise all the single particle states in a non-interacting system in one dimension with quenched random disorder~\cite{Anderson}. But it is feasible to see a localization to delocalization transition even in one dimension in models with deterministic quasiperiodic potentials, such as the Aubry-Andre (AA) model~\cite{AA} and models with Fibonacci potentials~\cite{Giamarchi}. 
These deterministic models have lately been investigated in the setting of many-body localization (MBL) both theoretically~\cite{Huse,Sdsarma,Subroto,khemani,sdsarma2019,RG_QP,Mirlin_Imb,Soumya,Yevgeny-AA,Piotr_AA,yp}  and experimentally~\cite{expt1,expt3,expt4,expt5,expt6} in the presence of interactions.  Despite active ongoing research in the field, the nature of the delocalization to MBL transition in these systems remains a mystery.

Finite-size scaling characteristics of the MBL transition in systems with AA potential have been studied previously using numerical exact diagonalization~\cite{khemani,Sheng,Piotr_AA,Yin}. Finite-size scaling of bipartite entanglement entropy and the standard deviation in entanglement entropy, under the assumption that the correlation length diverges as a power-law at the transition point $\xi \sim |h-h_c|^{-\nu}$, was performed to find the critical exponent $\nu < 1$ but close to one~\cite{khemani}. Later studies on finite-size scaling of entanglement entropy and sublattice fluctuations in magnetisation obtained $\nu =1.1$ for similar system sizes~\cite{Sheng}. In a more recent analysis of the finite-size scaling of the level spacing ratio across the MBL transition in interacting AA model a much smaller critical exponent of $\nu=0.54$ was obtained~\cite{Piotr_AA}. In complete contrast to these numerical studies, real space renormalization group calculation predicted a critical exponent $\nu =2.4$~\cite{RG_QP} and a similar exponent was obtained from the analysis based on local integral of motion for quasiperiodic MBL systems~\cite{LIOM_QP}. Thus, there is no consensus on the value of the correlation length critical exponent and the universality class to which the MBL transition in quasiperiodic systems belongs to.

A major issue in understanding the nature of the MBL transition in quasiperiodic systems is the lack of any generalized criterion like  Chayes-Chayes-Fisher-Spencer (CCFS)~\cite{CCFS} which is applicable for continuous transitions in systems with fully random disorder. For systems with quasiperiodic potential, CCFS criterion does not hold but there is an analogue of the relevance-irrelevance Harris-criterion~\cite{Harris} for quasiperiodic systems known as Luck criterion~\cite{Luck}. According to Luck criterion, a continuous transition in a clean d-dimensional system is stable with respect to the quasi-periodicity in the system if the correlation-length critical exponent $\nu > 1/d$~\cite{Luck}. But situations in which the transition is caused by quasi-periodicity itself and the clean system does not undergo any transition, relevance-irrelevance Luck criterion should not be applied. The MBL transition in quasiperiodic systems should, in principle, belong to a different universality class than the MBL transition in random disordered systems. Unlike the deterministic quasiperiodic potential, MBL systems with random disorder occasionally exhibit regions of very weak and large disorder. In systems with random disorder, these uncommon regions are known to be significant close to the MBL transition~\cite{RG_Ehud,RG_Potter,RG_Dumit,Huevneers,khemani_PRX} while these rare regions are absent in deterministic quasiperiodic systems. Second, while the delocalized side of systems with quasiperiodic potential has ballistic dynamics for the non-interacting case and super-diffusive dynamics for the interacting case~\cite{Dhar,Yevgeny-AA,yp_nee}, the delocalized side of systems with random disorder exhibits diffusive dynamics. MBL transition in quasiperiodic systems must therefore be distinct from MBL transition in random systems.

\begin{figure*}[ht]
  \begin{center}
    \vskip0.5cm
    \hspace{-1cm}
  \includegraphics[width=6.6in]
                  {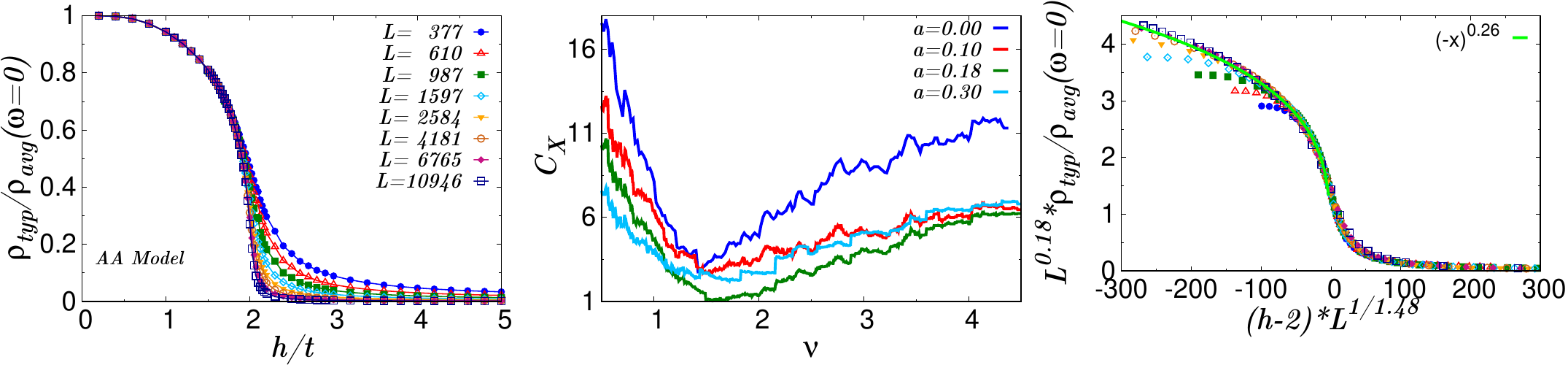}
  \caption{Non-Interacting AA Model - First panel shows the ratio of the typical to average LDOS $\rho_{typ}/\rho_{avg}$
   at $\omega=0$, as a function of the disorder strength $h$. The data presented has been obtained for $\eta$ equal to twice the average level spacing and has been averaged over $500-8000$ independent disorder configurations for $L=10946-377$ respectively. Second panel shows the cost function $C_X$ for $X=L^a[\rho_{typ}(\omega=0)/\rho_{avg}(\omega=0)]$ as a function of the correlation length exponent $\nu$ for $h_c=2t$ for various values of $a$. The minimum of the cost function occurs at $\nu \sim 1.48$ for $a=0.18$. Third panel shows the finite-size scaling of the ratio of the typical to average LDOS, $L^a[\rho_{typ}/\rho_{avg}]$, at $\omega=0$. A good scaling collapse is observed close to the transition point and even away from it on the localized side of the transition. Green curve shows the asymtotic behaviour of $\frac{\rho_{typ}}{\rho_{avg}} \sim (h_c-h)^{\beta}$ with $\beta = a\nu \sim 0.26$.   }
  \label{AA_nonint}
\vskip-1cm
\end{center}
\end{figure*}

In this work we study a one-dimensional system of fermions in the presence of AA potential~\cite{AA}. 
We examine the system in the limit of no interaction as well as in the presence of nearest neighbour repulsion. We mainly explore the universal properties of  single-particle excitations across the localization and MBL  transitions. To the best of our knowledge, single particle excitations have not been explored so far for systems with AA potentials. For the many-body interacting system we compute single-particle Green's function in real space in highly excited many-body eigenstates of the system and analyze the corresponding local density of states (LDOS).  
We demonstrate that the finite-size scaling of the ratio of the typical to average value of the LDOS yields correlation length critical exponent $\nu \ge 1$ for the interacting as well as non-interacting system. This should be compared with the finite-size scaling of single-particle LDOS for the MBL system with random disorder for which $\nu_{rand} \ge 2$ in consistency with the CCFS criterion~\cite{Atanu2}. Further for MBL systems with quasiperiodic potentials finite-size scaling of the level spacing ratio shows $\nu \sim 0.54$ which is consistent with earlier study on the AA model~\cite{Piotr_AA} and is also very close to the exponent obtained for MBL systems with random disorder~\cite{Piotr,Atanu2}. Additionally, the transition point obtained from level spacing ratio $h_{lsr}$ is smaller than that from the finite size scaling of LDOS $h_{c}$. This difference in transition points was also observed for MBL systems with random disorder~\cite{Atanu2} although the width $h_{c}-h_{lsr}$ of the intermediate phase is comparatively smaller for MBL systems with quasi-periodic potential. 

The rest of the paper is organised as follows. In Section II, we introduce the model explored in this work. In section III, we study single particle LDOS for the non-interacting model and perform finite-size scaling analysis using the cost function formalism.  In section IV we study single particle LDOS and level spacing ratio for the interacting AA model. We perform finite-size scaling for both the quantities. In section V we summarize our results and conclude with some remarks and open questions.

\section{II. Model}
We study a model of spin-less fermions in one-dimension described by the following Hamiltonian  
\be
H=-t\sum_{i}[c^\dagger_ic_{i+1}+h.c.] + \sum_i h_i n_i +\sum_{i} V n_in_{i+1}
\label{model}
\ee
with periodic boundary conditions. Here $t$ is the nearest neighbor hopping amplitude fixed to be one here, $V$ is the strength of nearest neighbour repulsion between Fermions and $h_i$ is the on-site potential of the form $h_i=h \cos(2\pi\beta i+\phi)$ where $\beta = \frac{\sqrt{5}-1}{2}$ is an irrational number and $\phi \in [0, 2\pi)$ is a random phase taken from a uniform distribution. We study this model at half-filling of fermions. In the non-interacting limit ($V=0$), $h_i$ corresponds to the Aubry-Andre potential~\cite{AA} which has a delocalization-localization transition at $h=2t$~\cite{AA,Hashomoto}. The model in \Eqn{model} has been studied extensively in the context of MBL, theoretically~\cite{Huse,khemani,Sheng,sdsarma2019,RG_QP,Mirlin_Imb,Soumya,Yevgeny-AA,Piotr_AA,Yin} as well as experimentally~\cite{expt1,expt2,expt3,expt4,expt5,expt6}.
 However, to the best of our knowledge, the analysis of the single-particle excitations and the corresponding LDOS to characterize the MBL phase and to understand the nature of the localization transition has not been explored so far for this model and this is the main focus of our work.
\section{III. LDOS for Non-Interacting quasiperiodic models}
In this section we discuss local density of states for the non-interacting AA  model of \Eqn{model}. The LDOS for the non-interacting model is given by $\rho_i(\omega)=\sum_n|\Psi_n(i)|^2\delta(\omega-E_n)$ where $E_n$ are the eigenvalues of the Hamiltonian in \Eqn{model} for $V=0$ and $\Psi_n(i)$ are the corresponding eigenfunctions. We introduce a small infinitesimal $\eta$ to broaden the
delta functions into Lorentzians such that $\rho_i(\omega)=\frac{1}{\pi}\frac{\eta|\Psi_n(i)|^2}{(\omega-E_n)^2+\eta^2}$. We chose $\eta$ to be a few times the average eigen value spacing. The typical value of LDOS $\rho_{typ}(\omega)$ is obtained by calculating the geometric average over the lattice sites, energy bin and various independent disorder configurations while the average value $\rho_{avg}(\omega)$ is obtained by simple arithmetic average over sites, energy bin and a large number of independent disorder configurations obtained by changing the random phase $\phi$ in \Eqn{model}. Fig.~\ref{AA_nonint} shows the ratio of typical to average value of the LDOS $\rho_{typ}(\omega)/\rho_{avg}(\omega)$ for $\omega\sim 0$ as a function of disorder strength $h$. For small values of $h$, $\rho_{typ} \sim \rho_{avg}$ and as $h$ increases, typical value reduces faster than the average value. For $h\ge 2$, $\rho_{typ}/\rho_{avg}$ is very small and shows a clear decrease as the chain size increases. Thus, single-particle excitations are suppressed on the localized side resulting in vanishingly small values of the typical LDOS while the excitation typically propagate over large length scale on the delocalized side. One sees a clear transition at $h=2$ as expected in this non-interacting AA model and $\rho_{typ}/\rho_{avg}$ acts as a good order parameter to distinguish the delocalized phase from the localized phase.

We perform finite-size scaling of $\rho_{typ}(\omega=0)/\rho_{avg}(\omega=0)$ assuming that the  characteristic length scale diverges with a power law $\xi \sim |h-h_c|^{-\nu}$ at the localization transition point $h_c$. As a result a normalized observable $X$ obeys the scaling $X[\delta,L] \sim \bar{X}(\delta L^{1/\nu})$ with $\delta=h-h_c$. For the non-interacting disordered system it is well known that $\rho_{typ}(\omega=0)/\rho_{avg}(\omega=0) \sim L^{-a}  \bar{X}(\delta L^{1/\nu})$ at the localization transition point; where $a$ is related to the position of peak in the singularity spectrum ~\cite{Janssen,Brillaux,Pixley}. To have a quantitative estimate of the scaling collapse, we calculate the cost-function for $X=L^a \rho_{typ}(\omega=0)/\rho_{avg}(\omega=0)$ as
\be
C_X=\frac{\sum_{j=1}^{N_{total}-1}|X_{j+1}-X_j|}{max\{X_j\}-min\{X_j\}} -1
\label{cost}
\ee
where $N_{total}$ is the total number of values of $\{X_i\}$ for various values of disorder $h$ and system sizes $L$ ~\cite{Prosen,Somen,Atanu2}. Each $X_i$ is a disorder averaged quantity over a large number of disorder realizations. One has to sort all $N_{total}$ values of $X_i$ according to increasing values of $(h-h_c)L^{1/\nu}$ which is then used in Eqn.~\ref{cost} to obtain the cost-function $C_X$. If $X$ is a smooth monotonic function of $(h-h_c)L^{1/\nu}$, then $\sum_{j=1}^{N_{total}-1}|X_{j+1}-X_j| = X_{N_{total}}-X_1 =max\{X_j\}-min\{X_j\}$. Thus, in case of an ideal collapse, $C_X \rightarrow 0$ but for a practical purpose here we look for a minimum of $C_X$ which is always positive.

For the non-interacting AA model, since the transition point $h_c$ is already known, we use the known value of the transition point, $h_c=2t$, and calculate the cost function for various values of the exponent $\nu$. Middle panel in Fig.~\ref{AA_nonint} shows the cost function $C_X$ vs the critical exponent $\nu$. The cost function has a minimum at $a=0.18$ and $\nu \sim 1.48 > 1$. The value of $\nu$ is consistent with earlier works on AA model which predicted $\nu \ge 1$ based on finite-size scaling of inverse participation ratio (IPR)~\cite{Hashomoto}. In the third panel of Fig.~\ref{AA_nonint} scaling collapse of $\rho_{typ}(\omega=0)/\rho_{avg}(\omega=0)$ is presented. A good quality of scaling collapse is obtained close to the transition point and even away from it on the localised side of the transition.  Ratio of typical to average DOS $\rho_{typ}(\omega=0)/\rho_{avg}(\omega=0)$ continuously goes to zero at $h_c=2t$ with the asymptotic form $(h_c-h)^\beta$ with $\beta=a\nu \sim 0.26$ as shown in the third panel of Fig.~\ref{AA_nonint}.

\begin{figure*}[ht]
  \begin{center}
    \vskip0.5cm
    \hspace{-1cm}
  \includegraphics[width=6.6in]
                  {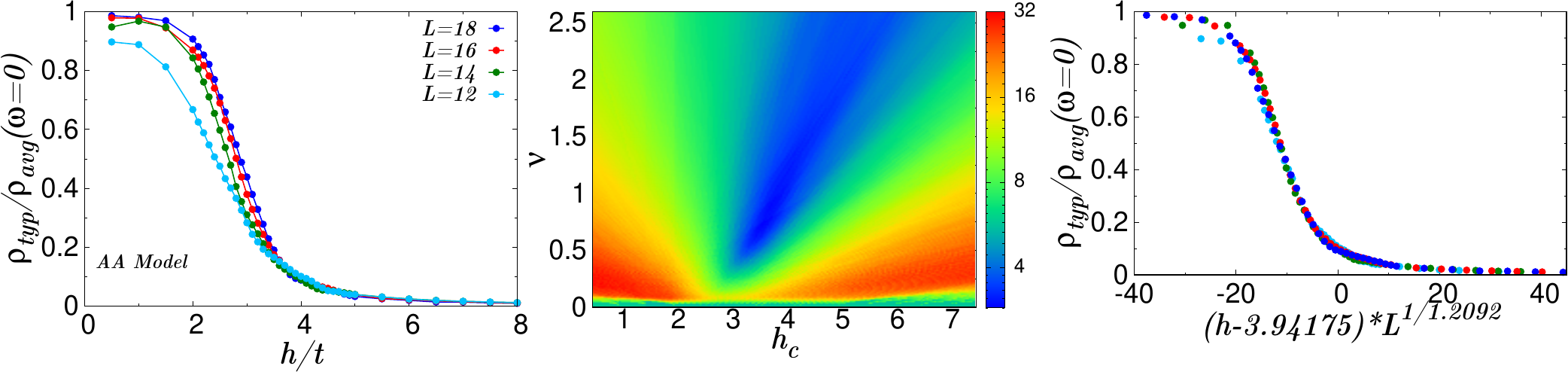}
  \caption{Interacting AA Model - First panel shows the ratio of the typical to average LDOS $\rho_{typ}/\rho_{avg}$
   at $\omega=0$, as a function of the disorder strength $h$. The data shown has been obtained from Green's function calculated in the middle of the many body spectra with $E\sim 0.5$ for $\eta = 0.01$ and has been averaged over $50-10000$ independent disorder configurations for $L=18-12$ respectively. Second panel shows the cost function $C_X$ for $X=\rho_{typ}(\omega=0)/\rho_{avg}(\omega=0)$ as a function of the correlation length exponent $\nu$ and the critical disorder strength $h_c$. Third panel shows the finite-size scaling of the ratio of the typical to average LDOS $\rho_{typ}/\rho_{avg}$ at $\omega=0$. We find $h_c \sim 3.94$ and $\nu \sim 1.21$ from minimisation of the cost function as shown in Fig~\ref{minm_AA}.}
  \label{AA_int}
\vskip-1cm
\end{center}
\end{figure*}

\section{IV. Nature of MBL transition in quasiperiodic systems}
In this section, we study single-particle Green's function in real space for interacting AA model in~\Eqn{model}. We will mainly focus on the Green's function calculated for the  eigenstates in the middle of the many-body spectrum. This is because MBL transition involves highly excited many-body eigenstates~\cite{Alet_rev, Abanin_rev,Huse_rev}. Many-body eigenstates in the middle of the spectrum get localized in the end as the disorder strength is increased for a fixed strength of interaction. Density of states for the model in \Eqn{model} is sharply peaked in the middle of the spectrum and hence an infinite temperature limit, which basically gives average over the entire spectrum, will also have a dominant contribution from states in the middle of the spectrum. Thus, we calculate single particle Green's function for $E=\frac{E_n-E_{min}}{E_{max}-E_{min}} \sim 0.5$.
The Green's function in the $nth$ eigenstate is defined as $ G_{n}(i,j, t) = -i \Theta(t) \langle \Psi_n  \vert \{ c_i(t), c_j^\dg(0)\} \vert \Psi_n \rangle$
where $i,j$ are lattice site indices. The Fourier transform of $G_n(i,j,t)$ in the Lehmann representation can be written  as
\be
G_n(i,i,\omega) = \sum_m \frac{|\la\Psi_m|c^\dagger_i|\Psi_n\ra|^2}{\omega+i\eta-E_m+E_n} +\frac{|\la\Psi_m|c_i|\Psi_n\ra|^2}{\omega+i\eta+E_m-E_n}
\label{Gn}
\ee
Here if $|\Psi_n\ra$ is the $nth$ eigenstate of the Hamiltonian in \Eqn{model} for $N_e$ particles in the chain, states $|\Psi_m\ra$ used in the 
first (second) terms in \Eqn{Gn} are obtained from the diagonalization of $N_e+1$ ($N_e-1$) particle systems. We would like to emphasize that unlike in the non-interacting case, for the interacting system there may not exist a general relation between the finite-size scaling of LDOS and inverse participation ratio (IPR) in the Fock space.  Note that though we are calculating $G_n(i,i,\omega)$ mainly for many-body eigenstates in the middle of the spectrum, the Lehmann sum in \Eqn{Gn} still runs over all values of $E_m$.  All the data shown for the interacting model is for $V=t$.
$\eta$ is a positive infinitesimal and is set to be a small finite value for sufficient broadening of the delta function into Lorentzian. Since many-body eigenstates in the MBL phase are multifractal, a careful choice of $\eta$ is required~\cite{Altshuler_LDOS}. Typical value of the LDOS depends on the broadening $\eta$. In the thermodynamic limit, in the localized phase the typical value of the LDOS scales proportionally to $\eta$ while in the delocalized phase the typical LDOS is independent of $\eta$. For a finite size system, this independence of typical LDOS in the delocalized phase is seen for a range of $\eta$ between the average value of the level spacing of a system of size $L$ and the average level spacing of the system of size equal to the correlation length~\cite{Mirlin_LDOS,Altshuler_LDOS,LDOS3}.  For a finite size system, we fix an $\eta$ to be larger than average level spacing  such that for small values of $h$, $\rho_{typ}$ is independent of $\eta$ and for very large values of $h$, $\rho_{typ}$ increases with $\eta$ for various system sizes under consideration.  This is consistent with the approach used in most of the numerical works e.g.~\cite{Parisi,Kubo1,Kubo2}. 
\begin{figure}[]
  \begin{center}
    \vskip0.5cm
  \includegraphics[width=3.5in]
                  {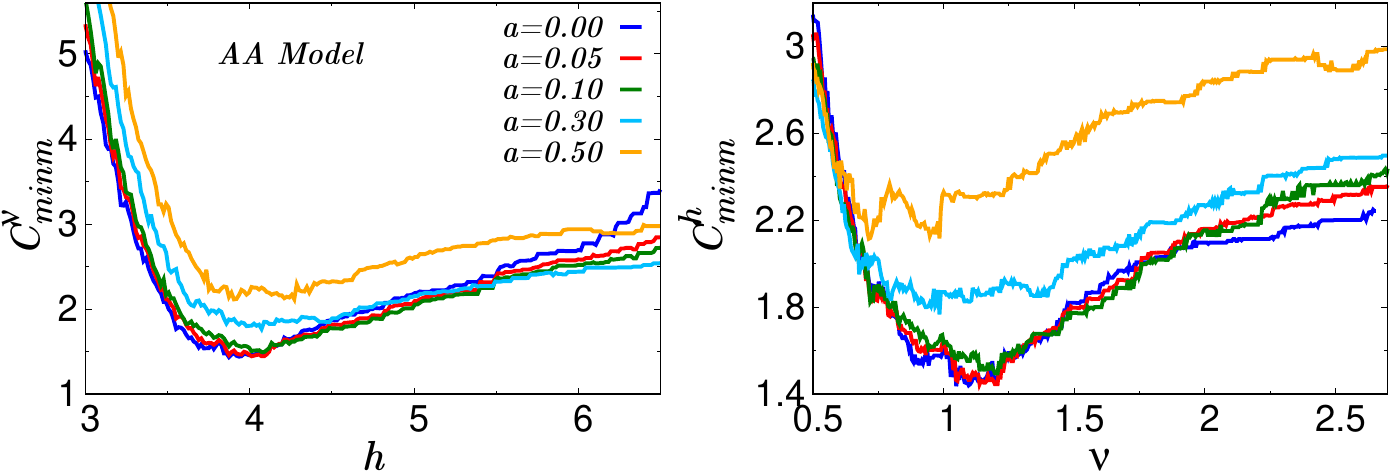}
  \caption{Minimum of the cost function $C_{minm}^\nu$ as a function of the disorder strength $h$ for various values of $a$ for $\rho_{typ}(\omega=0)/\rho_{avg}(\omega=0)$ for the interacting AA model. The right panel shows $C_{minm}^h$ as a function of the correlation length exponent $\nu$.}
  \label{minm_AA}
\vskip-1cm
\end{center}
\end{figure}

Fig.~\ref{AA_int} shows the ratio of the typical to average value of the LDOS for interacting AA model as a function of disorder strength $h$ for various system sizes. In close similarity to the non-interacting case, for weak disorder $\rho_{typ}(\omega=0) \sim \rho_{avg}(\omega=0)$  with the ratio being close to one. However, the ratio increases as the system size increases for small values of $h$ in contrast to the non-interacting weak disorder case. As the disorder strength increases, $\rho_{typ}(\omega=0)/\rho_{avg}(\omega=0)$ decreases and becomes vanishingly small in the MBL phase.

\begin{figure*}[]
  \begin{center}
    \vskip0.5cm
    \hspace{-1cm}
  \includegraphics[width=6.6in]
                  {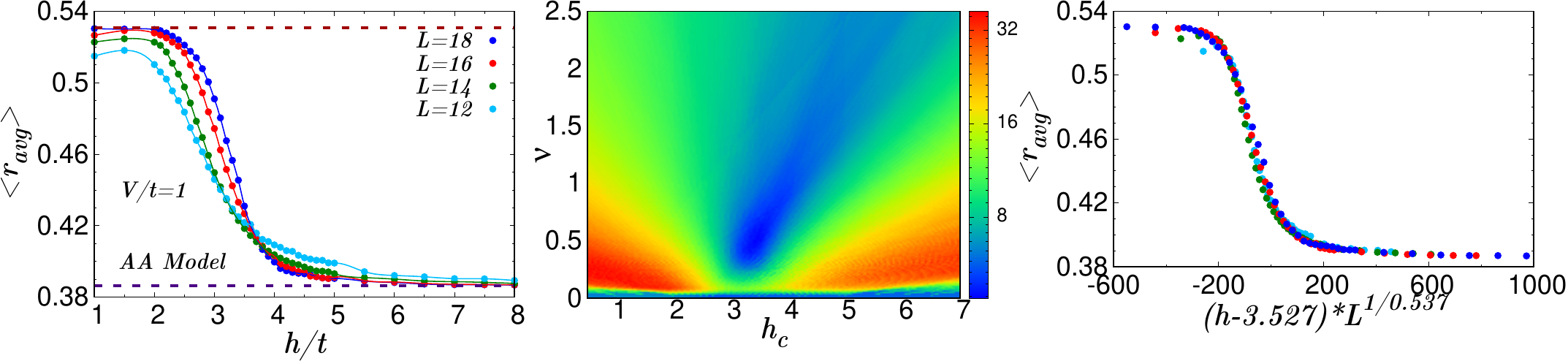}
  \caption{First panel shows the disorder-averaged level spacing ratio, $r_{avg}$, as a function of disorder strength $h$
   at $V/t=1$ for various system sizes for the interacting AA model. Data shown has been averaged over $50-10000$ disorder configurations for $L=18-12$ respectively. Second panel shows the cost function $C_X$ for $X=r_{avg}$ as a function of the correlation length exponent $\nu$ and the  disorder strength $h_c$. The minimum of the cost function occurs at $\nu \sim 0.537$ at $h_{lsr} \sim 3.527$. The third panel shows the finite-size scaling of the disorder-averaged level spacing ratio.}
  \label{lsr_AA}
\vskip-1cm
\end{center}
\end{figure*}

We further calculate the cost function defined in Eqn.~\ref{cost} in the $h-\nu$ plane as one needs to determine the transition point $h_c$ as well as the critical exponent $\nu$ for the interacting model. To determine the exact position of the transition point $h_c$ and the critical exponent at the transition point, the cost-function $C(h_c,\nu)$ is minimised w.r.t $\nu$ for each value of $h_c$ and various values of $a$. This results in $C_{min}^{\nu}$  which has been plotted as a function of $h_c$ in the left panels of Fig.~\ref{minm_AA}. The global minima w.r.t $h_c$ is obtained by finding the minima of $C_{min}^{\nu}$ as a function of $h_c$. Similarly, minimising the cost function w.r.t $h_c$ for each value of $\nu$ results in $C_{min}^h$ vs $\nu$ plot shown in the right panels. The global minima w.r.t $\nu$ is obtained by finding the minima of $C_{min}^h$ w.r.t $\nu$. As one can see from Fig.~\ref{minm_AA}, that the cost function has a minimum at $h_c \sim 3.94$, $a\sim 0$  and $\nu\sim 1.21$. Any non-zero value of $a$ increases the cost-function in most the parameter regime. With these values of $h_c$ and $\nu$ we obtain a very good quality scaling collapse of the data as shown in the third panel of Fig.~\ref{AA_int}, especially in the vicinity of the transition. This shows that even in the interacting AA model, critical exponent $\nu > 1$. Interestingly the value of $\nu$ obtained is very close to that obtained in earlier work from finite-size scaling of bipartitie entanglement entropy and sublattice magnetisation~\cite{Sheng}.

We also analyzed the finite-size scaling of the level spacing ratio averaged over the entire spectrum, which is frequently used to study the MBL transition. Fig.~\ref{lsr_AA} shows the level spacing ratio $r_n=\frac{min\{\Delta_n,\Delta_{n+1}\}}{max\{\Delta_n,\Delta_{n+1}\}}$ with $\Delta_n=E_{n+1}-E_n$ as a function of $h$ for various system sizes. The data shown has been obtained by averaging over the entire spectrum for a given disorder configuration and then averaged over a large number of independent disorder configurations.
As expected, for weak disorder $r_{avg}$ is close to the average value for Wigner-Dyson distribution and for strong disorder $r_{avg}$ is close to the average value for a Poisson distribution. The cost function calculated using Eqn.~\ref{cost} is shown in the middle panel of Fig.~\ref{lsr_AA}. 
\begin{figure}[]
  \begin{center}
    \vskip0.5cm
  \includegraphics[width=3.5in]
                  {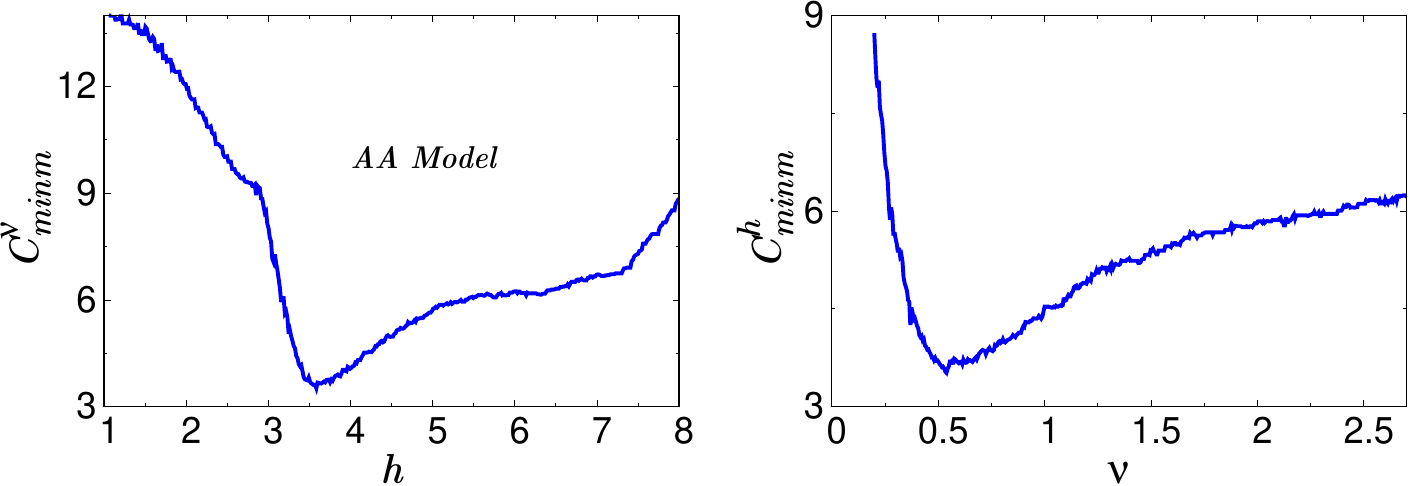}
  \caption{Interacting AA Model - Minimum of the cost function $C_{minm}^\nu$ for $r_{avg}$ as a function of the disorder strength $h$. The right panel shows $C_{minm}^h$ as a function of the correlation length exponent $\nu$.}
  \label{lsr_minmAA}
\vskip-1cm
\end{center}
\end{figure}
A proper minimisation procedure of the cost function, shows that the minimum occurs at $h_{lsr}=3.527$ and $\nu_{lsr}=0.537$, as shown in Fig.~\ref{lsr_minmAA}. This value of exponent $\nu$ is consistent with the recent work on AA model~\cite{Piotr_AA} and is also very close to the value obtained for the MBL systems with random disorder~\cite{Atanu2}. Further, the critical exponent obtained from the level spacing ratio is much smaller than that obtained from the finite-size scaling of LDOS.  We believe that, just like it has been argued for  MBL system with random disorder~\cite{Atanu2}, the level spacing ratio obeys dimensionality of the effective Anderson model in Fock space and hence might have critical exponent $\nu \ge 1/d_F$ where $d_F \sim L$ is the typical connectivity of the Fock space configurations resulting in exponents $\nu < 1$ for MBL systems with AA model as well as random disorder. Finite-size scaling of LDOS, on the other hand, provides very different values of the exponent for MBL systems with AA model and random disorder~\cite{Atanu2}.
It is important to notice that the scaling collapse for level spacing ratio using the parameters obtained from the minimization of the cost function is not as good as the collapse for the ratio of typical to average values of LDOS and may be improved by studying larger system sizes.

Another important observation to be made here is that the transtion point $h_c$ obtained from the finite-size scaling of LDOS is consistent with the predictions from local integrals of motion for the interacting AA model~\cite{LIOM_QP} and is slightly larger than the transition point obtained from level spacing ratio $h_{lsr}$. Studies on spin chains using time dependent variation principle have also found the transition point from density imbalance to be larger than the one obtained from level spacing ratio~\cite{Mirlin_Imb}. One possible explanation for this may be that the two quantities have different approach to the thermodynamic limit and the two transition points should come closer for bigger system sizes.
In MBL systems with random disorder, where $h_c-h_{lsr}$ is much larger~\cite{Atanu2} compared to what we have observed for the interacting AA model, rare region effects~\cite{RG_Ehud,RG_Potter,RG_Dumit,Huevneers,khemani_PRX} are supposed to be the reason behind this intermediate phase. This difference in the two transition points for the MBL systems with random disorder is also consistent with the scenario of system-wide rare resonances~\cite{avalanche}. But in deterministic model that we are studying here, there are no rare regions of disorder and hence what may be the mechanism behind an intermediate phase or a cascade of transitions for the interacting quasiperiodic models, is not obvious. Thus, we believe that the two transition points should merge as the system-size increases.

\section{V. Conclusions and Discussions}
In this work, we investigated the characteristics of the localization and MBL transition in deterministic quasiperiodic AA chains. In particular, we calculated the local density of states (LDOS) of single-particle excitations across the delocalization-localization transition in the interacting as well as non-interacting systems. For the interacting system, we analysed single-particle excitations generated in highly excited many-body eigenstates. We performed finite-size scaling of the ratio of the typical to average value of the LDOS, assuming that the characteristic length scale $\xi$ diverges as a power law at the transition point $\xi \sim |h-h_c|^{-\nu}$.  In the non-interacting as well as interacting quasiperiodic systems, we found the critical exponent $\nu \ge 1$ which is consistent with the IPR scaling for non-interacting AA model~\cite{Hashomoto} and the finite-size scaling of bipartite entanglement entropy for the interacting AA model~\cite{Sheng}.

We also studied level spacing ratio of consecutive eigenvalues for MBL systems with quasiperiodic AA potential. Finite-size scaling of level spacing ratio under the assumption of power-law divergence of the correlation length gives $\nu 0.54$. Interestingly, the value of $\nu$ for quasiperiodic system is very close to that obtained for the MBL system with random disorder~\cite{Piotr,Atanu2}. It is well known that $\nu$ resulting from level spacing ratio severely violates the CCFS criterion for MBL systems with random disorder. 
One of the potential causes of this might be that the level spacing ratio obeys dimensionality of the effective Anderson model in the Fock space~\cite{Atanu2} rather than the  physical dimension of the system. Also the transition point obtained from level spacing ratio $h_{lsr}$  is always smaller than $h_c$  obtained from LDOS for systems with quasiperiodic as well we random disorder. Whether the two transitions will merge in the thermodynamic limit needs to be explored.

Intuitively, one would expect the MBL transition in quasiperiodic systems to be in a different universality class than the MBL transition in systems with random disorder. MBL systems with random disorder have rare regions of very weak and large disorder while the deterministic quasiperiodic potential does not have those rare regions. The delocalized side of the random disorder systems have diffusive dynamics while the delocalized side of the systems with quasiperiodic potentials have super-diffusive dynamics. In complete contrast to this, from the finite-size scaling of the level spacing ratio, the MBL systems with random disorder and quasiperiodic potential seems to belong to the same universality class. Our finite-size scaling analysis of the LDOS reveals  a clear distinction between the MBL systems with quasiperiodic and random potentials. According to the finite-size scaling of LDOS $\nu\ge 1$ for MBL systems with quasiperiodic potential, whereas $\nu \ge 2$ for MBL systems with random disorder~\cite{Atanu2}. 
 It would be interesting to come up with more physical quantities which can support these findings and can help in understanding the nature of the MBL transition in quasiperiodic systems.

\section{Acknowledgements}
A.G. would like to acknowledge A. D. Mirlin for many insightful discussions and critical comments. A.G. would also like to acknowledge V. Ravi Chandra and Atanu Jana for discussions on related projects. Y. P. would like to acknowledge the National Research Foundation of Korea for the financial assistance.  We acknowledge National Supercomputing Mission (NSM) for providing computing resources of PARAM Shakti at IIT Kharagpur, which is implemented by C-DAC and supported by the Ministry of Electronics and Information Technology (MeitY) and Department of Science and Technology (DST), Government of India and SINP central cluster facilities.

\bibliography{bibliography}

\end{document}